\documentclass{PoS}
\usepackage{epsfig}

\title{Studies of $XYZ$ states at BESIII}

\ShortTitle{Studies of $XYZ$ states at BESIII}

\author{\speaker{Bin Wang}\thanks{Talk given at XXXIX International Conference on High Energy Physics (ICHEP2018), 4 July - 11, 2018, Seoul - Korea}\\
        Institute of High Energy Physics, Beijing 100049, China\\
        E-mail: \email{wangbin@ihep.ac.cn}}

\abstract{
With about 12 fb$^{-1}$ of collected data useful for the study of $XYZ$ states,
BESIII collaboration continues the exploration of these exotic charmoniumlike states.
Recent results of the measurements of the line-shape of $e^+e^-\to\pi^0\pi^0\psi(3686)$, $KKJ/\psi$, and $\pi^+D^0D^{*-}$,
as well as the $J^P$ determination of $Z_c(3900)$ and $Z_c(3900)$ observed in $e^+e^-\to\phi\chi_{c1, 2}$ at
$\sqrt{s}=4.6$~GeV will be presented.
}

\FullConference{ICHEP 2018, XXXIX International Conference on High Energy Physics\\
		4-11 July 2018\\
		Seoul, Korea}

\begin{document}

\section{Introduction}

Charmonium states are composed of one charmed quark and one anti-charmed quark ($c\bar{c}$).
States below the charm quark pair threshold in mass (2$m_D$) are well described by the potential mode~\cite{PRD32-189},
while many states above 2$m_D$ have not been observed yet.
In addition, there are many unexpected states which are called charmoniumlike or $XYZ$ states
and have a charmonium final states but no conventional charmonium states assignment
observed in recent years~\cite{pdg}.
Searching for the new decay modes of known charmonium or charmoniumlike states
and new charmonium-like states is helpful for the interpretation of these charmonium-like states.

BESIII has collected large data samples of electron positron collisions with center-of-mass (c.m.)
energy between 3.8 and 4.6~GeV, the total integrated luminosity is about 12~fb$^{-1}$.
Recent results of the measurements of the $XYZ$ physics will be described in the following sections.

\section{Determination of $J^P$ of $Z_c(3900)$}

The charged charmoniumlike state $Z_c(3900)$ was observed in the process $e^+e^-\to\pi^+\pi^-J/\psi$
by the BESIII~\cite{PRL110-252001} and Belle~\cite{PRL110-252002} and confirmed by the CLEO-c~\cite{PLB727-366-2013}.
As the first $Z_c$ state observed by more than one experiment, it is composed of at least four quarks.
Many theoretical interpretations of its nature and decay dynamics
have been put forward~\cite{CSB-59-3815, PRD88-054007}.
BESIII has also reported the observations of the neutral $Z_c(3900)$ in the process
$e^+e^-\to\pi^0\pi^0J/\psi$~\cite{PRL115-112003},
$Z_c(4020)$ in $e^+e^-\to\pi^{+, 0}\pi^{-, 0}h_c$~\cite{PRL111-242001, PRL113-212002},
$Z_c(4025)$ in $e^+e^-\to\pi^{\pm, 0}(D^*\bar{D}^*)^{\mp, 0}$~\cite{PRL112-132001, PRL115-182002},
and $Z_c(3885)$ in $e^+e^-\to\pi^{\pm, 0}(D\bar{D}^*)^{\mp, 0}$~\cite{PRL112-022001, PRD92-092006, PRL115-222002}.
Are $Z_c(3900)$ and  $Z_c(3885)$ the same state and do they have the same quantum number assignment?
The experimental determination is one of the most important expectations of theorists.

Recently, BESIII reported the determination of the $J^P$ of $Z_c(3900)$~\cite{PRL119-072001},
which is $J^P = 1^+$ with a statistical significance larger than 7$\sigma$ over other quantum number assumptions,
in the partial wave analysis of the process $e^+e^-\to\pi^+\pi^-J/\psi$ using 1.92~fb$^{-1}$ data samples at $\sqrt{s} = 4.23$ and 4.26~GeV.
There are six contributions
($e^+e^-\to\sigma J/\psi$, $f_0 J/\psi$, $f_0(1370) J/\psi$, $f_2(1270) J/\psi$, $Z_c^{\pm}\pi^{\mp}$, and non-resonant processes)
and five assumptions of the $J^P$ of $Z_c(3900)$ ($1^+$, $0^-$, $1^-$, $2^+$, and $2^-$) are considered in the fit.
Using a simultaneous fit for the data samples at 4.23 GeV and 4.26 GeV,
where the $Z_c(3900)$ state is described by a Flatte-like formula,
the mass and coupling parameters ($g_1^\prime$ and $g_2^\prime/g_1^\prime$) of $Z_c(3900)$ are measured to be (3901.5$\pm$2.7$\pm$38.0) MeV/c$^2$,
0.075 $\pm$ 0.006 $\pm$ 0.025)~GeV$^2$, and (27.1$\pm$2.0$\pm$1.9), respectively.
The fitted coupling constants are consistent with the measured decay width ration of $(D\bar{D}^*)^{\pm}$ and $\pi^{\pm}J/\psi$ final states.

\section{$e^+e^-\to\pi^0\pi^0\psi(3686)$}
Recently, BESIII reported the precise measurement of the cross sections for the processes
$e^+e^-\to\pi^+\pi^-J/\psi$~\cite{PRL118-092001} and indicated that the $Y(4260)$ resonances actually consists of two structures.
Two resonances, $Y(4220)$ and $Y(4390)$, are observed in the process $e^+e^-\to\pi^+\pi^-h_c$~\cite{PRL118-092002}.
BESIII has also reported more studies of these $Y$ states for understanding the puzzles with these states.
The $Y(4360)$ was first observed in $e^+e^-\to\gamma_{ISR}\pi^+\pi^-\psi(3686)$ by $BABAR$~\cite{PRL98-212001}
and subsequently confirmed by Belle~\cite{PRL99-142002} and BESII~\cite{PRD96-032004}.
By analogy, BESIII also reported the measurement of its neutral isospin $e^+e^-\to\pi^0\pi^0\psi(3686)$~\cite{PRD97-052001}.
The measured Born cross sections of $e^+e^-\to\pi^0\pi^0\psi(3686)$ are consistent with
those of $e^+e^-\to\pi^+\pi^-\psi(3686)$ from isospin symmetry.
In addition, a neutral charmoniumlike structure is observed in $\pi^0\psi(3686)$ with a mass of (4038.7$\pm$6.5)~MeV/$c^2$ at $\sqrt{s} = $4.416~GeV,
which confirms the structure in the charged mode. No significant $Z_c(3900)^0$ state is observed in the fit.

\section{$e^+e^-\to K\bar{K}J/\psi$}

As the first observed vector charmoniumlike state, the nature of the $Y(4260)$ is still unclear.
Besides of $e^+e^-\to\pi^+\pi^-J/\psi$, the $Y(4260)$ has also been searched for many modes in recent years,
such as $\pi^+\pi^-h_c$, $\omega\chi_{cJ}$, $\eta J/\psi$ and so on.
Especially, measuring the ratio of $e^+e^-\to K\bar{K}J/\psi$ and $e^+e^-\to\pi\pi J/\psi$ cross sections
provides a new insight into the nature of $Y(4260)$.
Recently, using 4.7~fb$^{-1}$ data sample from 4.189 to 4.600~GeV, BESIII reported the measurements of
the process of $e^+e^-\to K\bar{K}J/\psi$~\cite{PRD97-071101}.
The results show that the energy dependence of the cross section for $e^+e^-\to K^+K^-J/\psi$ is not
consistent with those of  $e^+e^-\to\pi^+\pi^-J/\psi$ in the region around 4.26~GeV.
The ratio of cross sections for $e^+e^-\to K^+K^-J/\psi$ and $e^+e^-\to K_S^0K_S^0J/\psi$ are consistent with expectations from isospin conservation. In addition, there is an evidence for a structure around 4.5~GeV
in the $e^+e^-\to K^+K^-J/\psi$ cross section that not present in the  $e^+e^-\to\pi^+\pi^-J/\psi$.

\section{$e^+e^-\to\pi^+D^0D^{*-}$}

BESIII also reported a precise cross section measurement of $e^+e^-\to\pi^+D^0D^{*-}$ process~\cite{piDDst}.
Two resonant structures are significant observed in this final states.
It indicates an evidence for open-charm production associated with the $Y$ states is observed for the first time.
The parameters of these two resonances are consistent with those measured in $e^+e^- \rightarrow \pi^{+}\pi^{-}h_c$. The first resonance are also consistent with those measured in
$e^+e^- \rightarrow \omega\chi_{c0}$~\cite{PRL114-092003} and $e^+e^- \rightarrow \pi^{+}\pi^{-}J/\psi$.
The first resonance is consistent with some of the theoretical calculations for the mass of $Y(4260)$ when explaining it as a $D\bar{D}_1(2420)$ molecule~\cite{PRD90-074039}.

\section{$e^+e^-\to\phi\chi_{c1,2}$}

BESIII has reported the cross section of $e^+e^- \rightarrow \omega\chi_{c0}$ and observed an intermediate resonance around 4226 MeV~\cite{PRL114-092003}.
Considering that $\omega$ and $\phi$ have the same spin, parity, and isospin,
the $\omega\chi_{cJ}$ and $\phi\chi_{cJ}$ should have a similar production mechanism,
so we study the $e^+e^-\to\phi\chi_{cJ}$.
Using 567~fb$^{-1}$ data sample at $\sqrt{s} = $ 4.60~GeV,
BESIII first observed the processes of $e^+e^-\to\phi\chi_{c1}$ and $\phi\chi_{c2}$~\cite{PRD97-032008}.
The corresponding Born cross sections are $(4.2^{+1.7}_{-1.0}\pm0.3)$ and $(6.7^{+3.4}_{-1.7}\pm0.5)$~pb,
respectively. No significant  $e^+e^-\to\phi\chi_{c0}$ and  $e^+e^-\to\gamma X(4140)$ were observed in this final states.

\section{Summary}

Recent studies of $XYZ$ states at BESIII collaboration are presented using large luminosity data samples collected above 4~GeV. 
BESIII is an active and successful experiment for the charmonium spectroscopy study.
In the future, BESIII will continue to take data for studying these $XYZ$ states more precisely
and increase the beam energy for studying the higher $XYZ$ states, such as $Y(4660)$.

\end{document}